\documentclass[journal=cmatex,manuscript=article]{achemso}
\usepackage[version=3]{mhchem} % Formula subscripts using \ce{}
\usepackage{pslatex}
\usepackage{xr}
\usepackage{gensymb} % for \degree symbol
\usepackage{textcomp} % for \textdegree symbol (better?)

\usepackage[russian, english]{babel} % needed to get the citations in Russian (Cyrrilic characters) to render properly
\graphicspath{{./Figures/}} % this places all graphics in the figures subdirectory
\title{Ball milling enables phase-pure synthesis of a temperature sensitive ternary chloride, \ce{MgZrCl6}}

\author{Christopher L. Rom}
\affiliation{National Renewable Energy Laboratory, Golden, Colorado 80401, United States}
\email{christopher.rom@nrel.gov}
\author{Austin Shotwell}
\affiliation{Department of Chemistry, Colorado School of Mines, Golden, Colorado 80401, United States}
\author{Sinclair Combs}
\affiliation{Department of Chemistry, Colorado School of Mines, Golden, Colorado 80401, United States}
\author{Autumn Peters}
\affiliation{Department of Chemistry, Colorado State University, Fort Collins, Colorado 80523, United States}
\author{Lauren Borgia}
\affiliation{Department of Chemistry, Colorado State University, Fort Collins, Colorado 80523, United States}
\author{James R. Neilson}
\affiliation{Department of Chemistry, Colorado State University, Fort Collins, Colorado 80523, United States}
\affiliation{School of Materials Science \& Engineering, Colorado State University, Fort Collins, Colorado 80523, United States}
\author{Annalise E. Maughan}
\affiliation{National Renewable Energy Laboratory, Golden, Colorado 80401, United States}
\alsoaffiliation{Department of Chemistry, Colorado School of Mines, Golden, Colorado 80401, United States}
\email{amaughan@mines.edu}

\begin{document}

\maketitle

\begin{abstract}
Ball milling is a powerful synthetic tool for discovering new inorganic materials. 
Inspired by the high ionic conductivity in \ce{Li2ZrCl6} synthesized via mechanochemistry, we synthesized \ce{MgZrCl6} with a similar method. 
High resolution synchrotron X-ray diffraction shows that \ce{MgZrCl6} is poorly crystalline after ball milling, but crystallizes in a layered hexagonal structure ($P\overline{3}1c$) after heat treatment. 
\textit{In situ} synchrotron X-ray diffraction reveals a narrow temperature window around 400 °C in which crystallization occurs. 
At higher temperatures, the phase decomposes.
Pair distribution function analysis shows 2D sheets of \ce{MgZrCl6} form after milling, with heating driving 3D crystallization. Raman spectroscopy also shows evidence of \ce{MgZrCl6} after milling.
Electrochemical impedance spectroscopy does not reveal ionic conductivity in \ce{MgZrCl6} (limit of detection ca. $1.4\times10^{-8}$ S/cm). 
In addition to supporting existing design rules for Mg-based solid electrolytes, this work shows the power of ball milling to synthesize temperature-sensitive inorganic compounds with high yield.
\end{abstract}

\section*{Introduction}

While traditional solid-state synthesis techniques rely on high temperatures to drive solid-state diffusion, ball milling can drive inter-diffusion between solids at low temperatures. 
Therefore, the mechanochemical technique can be an atomically efficient process for synthesizing materials ranging from halide perovskites for optoelectronic applications\cite{palazon2020making, 
%barbosa2024greenCs2AgSbCl6, 
ceriotti2024mechanochemicalKCuF3} to solid electrolytes for batteries.
Leading Li- and Na-based solid electrolytes, such as sulfide argyrodites,\cite{boulineau2012mechanochemical} ternary chlorides,\cite{kwak2021newLi2ZrCl6} and oxychlorides,\cite{zhao2024sodium} have been synthesized via ball milling.
However, solid-state Mg electrolytes remain a nascent research topic\cite{
% blazquez2023practicalPerspective, 
jaschin2020materialsReviewMg} 
which ball milling may expand.

Chlorides have emerged as a promising class of materials for \ce{Li^{+}} solid electrolytes owing to their high ionic conductivity, low electronic conductivity, and high oxidative stability,\cite{combs2022editors} but are underexplored for \ce{Mg^{2+}} conductivity.
To the best of our knowledge, only \ce{Mg$M_2X_8$} phases ($M$ = Al, Ga; $X$ = Cl, Br) have been studied as \ce{Mg^{2+}} ion conductors.\cite{tomita2020synthesisMgM2X8, tomita2021synthesisMgAl2Cl8} 
They were synthesized via high-temperature solid-state techniques and exhibit ionic conductivity ca. $10^{-6}$ to $10^{-5}$~{S/cm} at 127~°C.
Although these few examples under perform leading selenide\cite{canepa2017highMgMobility} and borohydride\cite{roedern2017magnesiumBorohydride} materials, further exploration of chlorides is warranted.
Solid state synthesis methods yielded chlorides with low-to-moderate \ce{Li+} ion conductors in the 1990's, but the ball-milling synthesis of \ce{Li3YCl6} produced a fast ion conductor in 2018 and started a renaissance of chloride-based solid electrolytes.\cite{asano2018solidLi3YCl6, combs2022editors}

\begin{figure}[ht!]
    \centering
    \includegraphics[width = \textwidth]{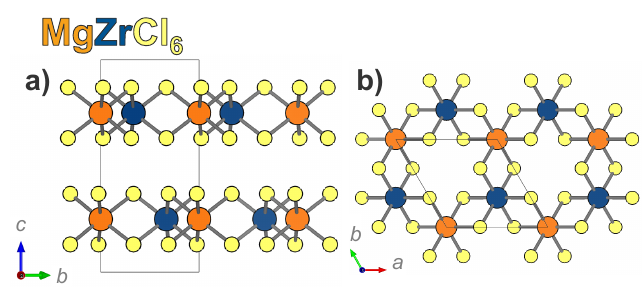} %3.2in
    \caption{a) View of the \ce{MgZrCl6} structure down the $a$ axis.  b) View of one layer of the structure looking down the $c$ axis. 
    }
    \label{fig:structure}
    % Source: MgZrCl6_figureMapping.pptx
\end{figure}
\ce{MgZrCl6} has a layered structure that may be conducive to \ce{Mg^{2+}} mobility (Figure \ref{fig:structure}), but it has not been studied for this property. We first noticed this phase as an intermediate in the metathesis reaction: \ce{2Mg2NCl + ZrCl4 -> MgZrN2 + 3 MgCl2}.\cite{rom2021bulk} A 2014 report by Salyulev and Vovkotrub noted that \ce{MgZrCl6} was previously studied to better understand corrosion in chloride-based metallurgical processes,\cite{russian2014_melts} and they referenced synthesis literature for these phases from the 1990's.\cite{salyulev1998raman, russian1990_MgZrCl6} We have not been able to access these original synthesis reports, but the 2014 report suggests that \ce{MgZrCl6} was synthesized using elevated \ce{ZrCl4} vapor pressures (ca.\ 22–59~atm) and in narrow temperature ranges (ca.\ 450-500~°C).\cite{russian2014_melts} 
Given the volatility of \ce{ZrCl4} (sublimation point, 331 °C),
we hypothesized that mechanochemistry may provide a route to phase pure \ce{MgZrCl6}.

\section*{Results and Discussion}
\begin{figure}[ht!]
    \centering
    \includegraphics[width = \textwidth]{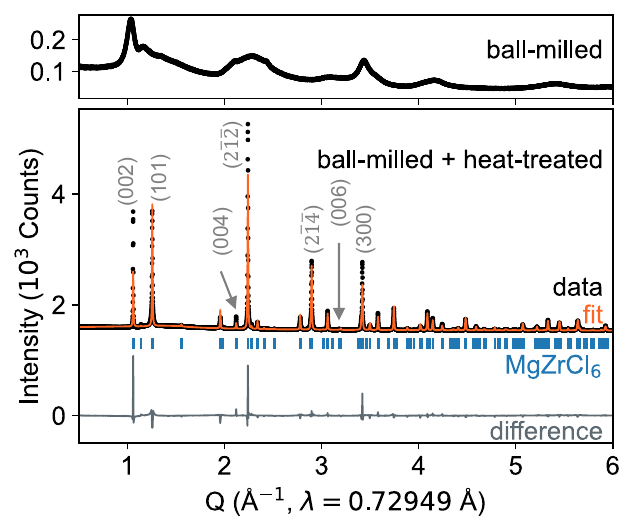}
    \caption{SPXRD of \ce{MgZrCl6} prepared by ball milling along with SPXRD of the sample after heat treatment at 350~°C for 2 h.
    % Traces are vertically offset for clarity. 
    }
    \label{fig:main_XRD}
    % Source: C:\Users\topas\Documents\ChrisRom\MgZrCl6_fitting_highResXRD_SSRL_NSLS-II\SSRL\MgZrCl6_fits.pro
    % See also:  % Source: MgZrCl6_figureMapping.pptx
\end{figure}
High resolution synchrotron powder X-ray diffraction (SPXRD) shows that high-energy ball milling (BM) of \ce{MgCl2 + ZrCl4} produced poorly crystalline \ce{MgZrCl6} (Figure \ref{fig:main_XRD}). Subsequent heat treatment (BM+HT) crystallizes \ce{MgZrCl6} in space group $P\bar{3}1c$ ($a = 6.35975(3)$~\AA{} and $c = 11.8428(1)$~\AA{}), isostructural with \ce{FeZrCl6} (ICSD Col. Code 39666).\cite{troyanov1992crystalFeZrCl6} 
Ball milling was crucial for the synthesis of phase pure \ce{MgZrCl6}, as hand-ground mixtures of reagents only reacted partially (Figures S1, S2).%\ref{fig:mass_loss}).

The crystal structure of \ce{MgZrCl6} (BM+HT) consists of layers stacked along the \textit{c} direction, with each layer containing edge-sharing [\ce{MgCl6}] and [\ce{ZrCl6}] octahedra. 
Within the layer, 2/3 of the octahedral sites are occupied with an alternating pattern of \ce{Mg^{2+}} and \ce{Zr^{4+}}, while the remaining 1/3 of octahedral sites are vacant. Consequently, each \ce{Mg^{2+}} is neighbored by three \ce{Zr^{4+}} and three vacant octahedra. 
The chloride anions form a hexagonal close-packed arrangement with a van der Waals gap between the layers. 
We note three significant peaks in the difference trace, indicating that the (002), ($2\bar{1}2$), and (300) Bragg peaks are under-fit by our model. 
Our attempts to improve our model with stacking faults, anisotropic peak broadening, and cation-disorder, were unsuccessful. We also considered a structural model based on \ce{TlYbI6} (ICSD Col. Code 138835), which also crystallizes in the $P\bar{3}1c$ space group but with different atomic coordinates: that model was substantially worse. It is possible for cations to disorder into van der Waals gap,\cite{mandujano2024kinetic} but our attempts to refine electron density in inter-layer sites did not substantially improve the fit. Single crystal diffraction measurements may be needed to more precisely determine the structure.

\begin{figure}[h]
    \centering
    \includegraphics[width = 0.7\textwidth]{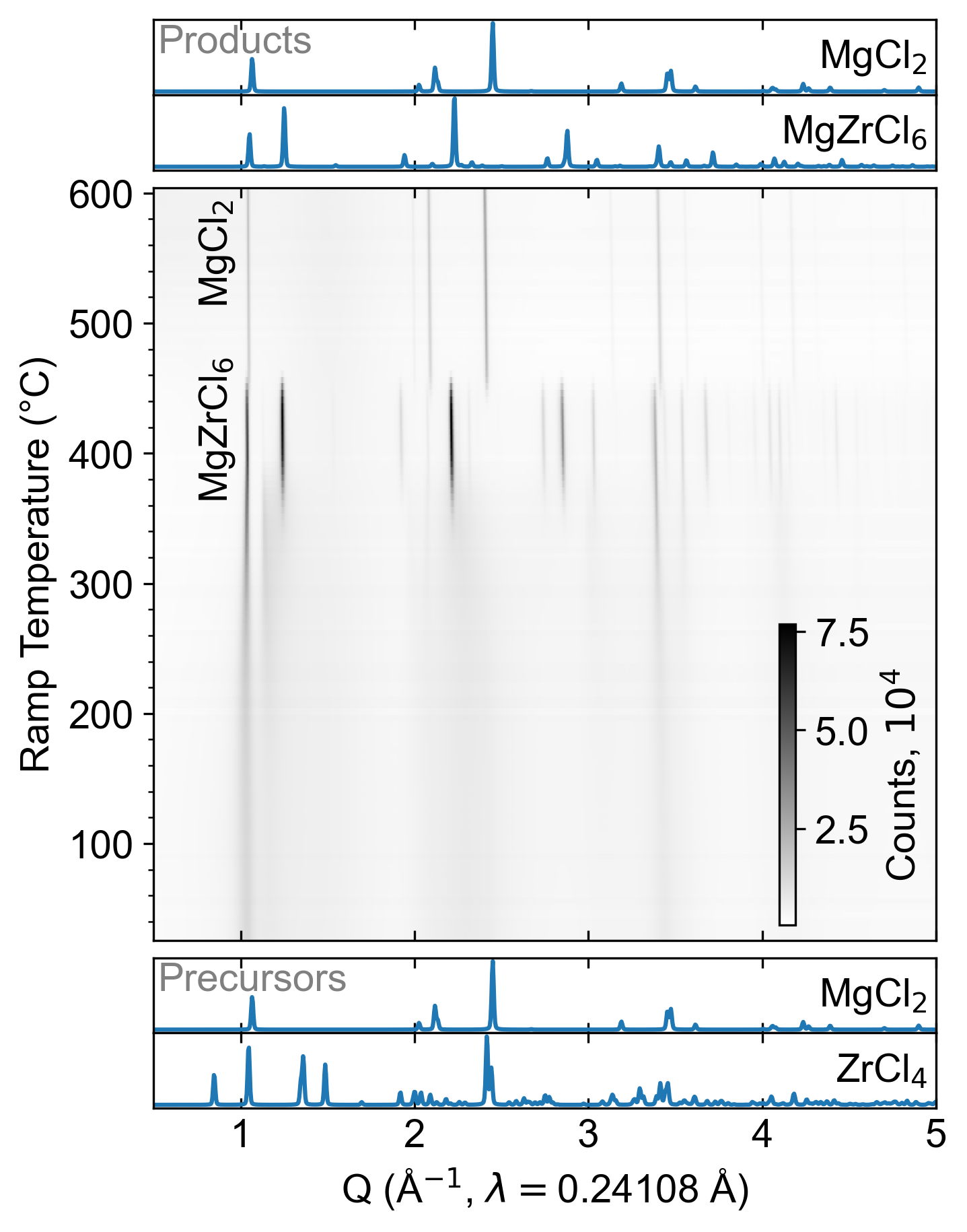}
    \caption{\textit{In situ} SPXRD of a BM mixture of \ce{MgCl2 + ZrCl4} upon heating at +10~°C/min. Simulated reference patterns for the precursors and products are shown at the bottom and top, respectively.
}
% Source: 
% exported from: http://localhost:8888/notebooks/Documents/Beamtime_2023_03/MgZrCl6%20in%20situ.ipynb
%Annotations added in MgZrCl6_figureMapping.pptx
% Greyscale option direct export from ipynb: /Users/crom/Documents/Beamtime_2023_03  
    \label{fig:main_inSitu}
\end{figure}
\textit{In situ} SPXRD shows the crystallization and decomposition pathway for \ce{MgZrCl6} from ball-milled precursors (Figure \ref{fig:main_inSitu}). Broad peaks are present at room temperature, indicating that ball-milling resulted in a poorly crystalline ternary phase. 
At 340~°C, the broad peaks of the initial phase sharpen, and additional reflections appear as \ce{MgZrCl6} rapidly crystallizes. 
At 460~°C, the \ce{MgZrCl6} peaks abruptly disappear, leaving behind only \ce{MgCl2}. 
This change shows that \ce{MgZrCl6} has limited thermal stability, decomposing to \ce{MgCl2} (s) and \ce{ZrCl4} (g) at moderate temperatures. 
These \textit{in situ} findings are consistent with our \textit{ex situ} results (Figure S1).%\ref{fig:mass_loss}). 

\begin{figure}[h]
    \centering
    \includegraphics[width = 0.7\textwidth]{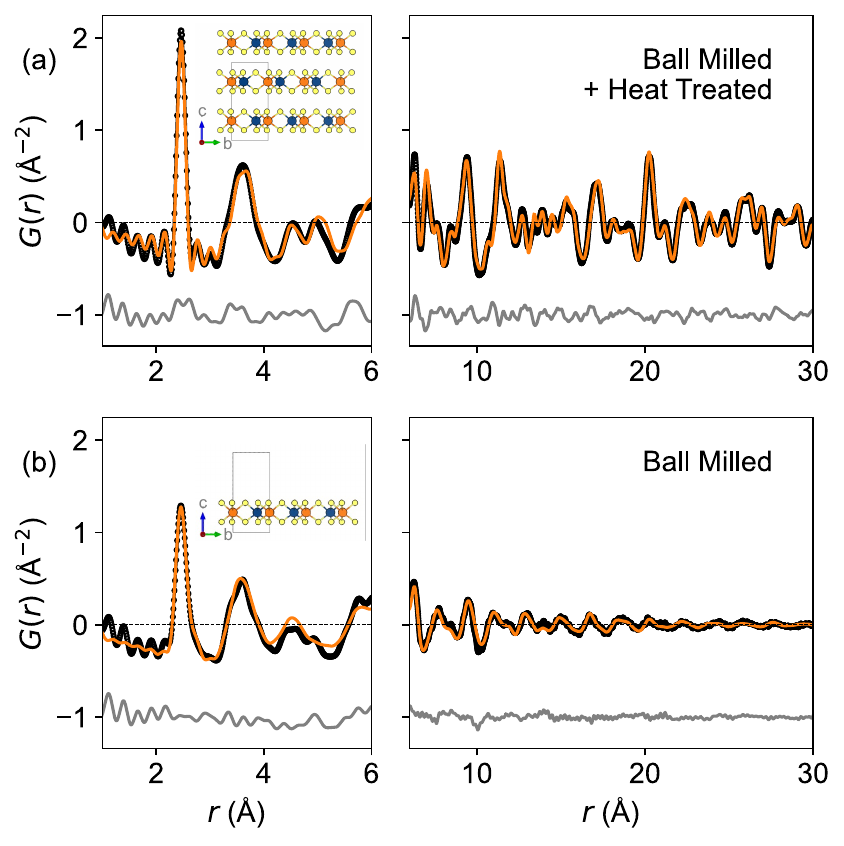}
    \caption{PDF of X-ray total scattering data from (a) BM+HT samples and (b) BM of \ce{MgZrCl6}. Data in black, fit in orange, difference in gray (offset vertically by 1 Å$^{-2}$). Insets show structural models.}
    \label{fig:PDF}
    %%% see PDFplotting/PlottingFits_MgZrCl6.ipynb
    % and MgZrCl6_figureMapping.pptx
\end{figure}

Pair distribution function (PDF) analysis of X-ray total scattering data suggest that BM \ce{MgZrCl6} has a similar local structure to BM+HT \ce{MgZrCl6}(Figure \ref{fig:PDF}). The BM+HT sample exhibits short and long range pair correlations that are well fit by the $P\bar31c$ \ce{MgZrCl6} model (Figure \ref{fig:PDF}a).  The PDF of the ball milled sample reveals significantly attenuated pair correlations beyond $r\approx6$~\r{A}.  Furthermore, the bulk crystal structure does not fit these data well beyond $r\approx$4 \r{A}, particularly at distances corresponding to the interlayer separations (Figure~\ref{fig:PDFmodelcmpr}). Instead, a composite model with a single layer of \ce{MgZrCl6} as implemented in {\sc PDFgui}\cite{Farrow_2007} following  Ref.~\citenum{Chen:iu5034} with a spherical truncation diameter of 50(30) \r{A}, provides the best fit to the data (Figure \ref{fig:PDF}b). A similar result was observed via PDF for the initial stages of growth FeS  from solution.\cite{beauvais2021resolvingFeS} This suggests that ball milling induces formation of \ce{MgZrCl6} sheets with octahedral coordination and some Mg-Zr ordering, but annealing is necessary to induce extended ordering.

\begin{figure}[h]
    \centering
    \includegraphics[width = 3.2 in]{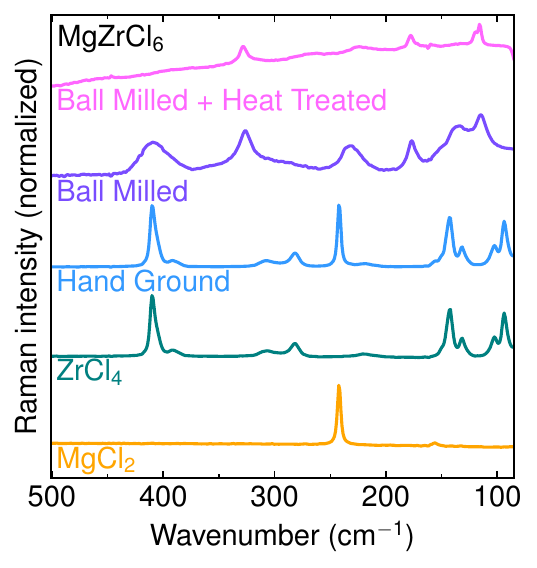}
    \caption{Background-subtracted Raman spectra of the crystallized \ce{MgZrCl6} (BM+HT) and poorly-crystalline ball milled \ce{MgZrCl6} compared with the handground precursor mix \ce{MgCl2 + ZrCl4} and the binary precursors. Raw spectra are shown in Figure S4.}
    \label{fig:main_raman}
    %Source; Austin
\end{figure}

Raman spectroscopy shows that the BM \ce{MgZrCl6} has structural motifs that are conserved upon crystallization (Figure \ref{fig:main_raman}). 
The Raman spectrum of BM+HT \ce{MgZrCl6} has peaks at 327 cm$^{-1}$, 177 cm$^{-1}$, and 116 cm$^{-1}$. 
Similar peaks also appear in the spectrum for the BM \ce{MgZrCl6}. 
While we do not precisely assign these vibrational modes, these shared peaks suggest that structural motifs of the crystallized \ce{MgZrCl6} are already present in the poorly crystalline material produced by the ball milling step, consistent with PDF analysis. 
The BM \ce{MgZrCl6} spectrum also has broad peaks at 410 cm$^{-1}$, 232 cm$^{-1}$, and 135 cm$^{-1}$ that roughly correspond to peaks from the binary halide precursors, suggesting that mechanochemical conversion to \ce{MgZrCl6} is incomplete (10 h milling time). 
In contrast, the spectrum from the handground sample of \ce{MgCl2 + ZrCl4} is merely a linear combination of the precursor spectra. These data show that ball milling initiates formation of the \ce{MgZrCl6} phase, which crystallizes on heating.

Given the open framework of \ce{MgZrCl6} with an ordered arrangement of vacant octahedra within layers and a van der Waals gap between layers (Figure \ref{fig:structure}), we hypothesized that \ce{Mg^{2+}} may be mobile in the structure. We performed AC electrochemical impedance spectroscopy (EIS) in a two-electrode configuration up to 95~°C.
Measurements on BM+HT \ce{MgZrCl6} did not exhibit charge transport behavior (Figure \ref{fig:main_EIS}). The poorly crystalline BM \ce{MgZrCl6} also did not exhibit \ce{Mg^{2+}} conductivity.
Rather, the materials show capacitive behavior consistent with a dielectric. 
Given the limit of detection for the measurement (approximately 4~M$\Omega$) along with the pellet dimensions (0.71 mm thick, 1.27 cm$^2$ cross-sectional area), we can rule out ionic conductivity above approximately $1.4\times10^{-8}$ S/cm. 
We attempted aliovalent substitution of Nb$^{5+}$ into \ce{MgZrCl6} in hopes of boosting ionic conductivity (Figure \ref{fig:EIS_supp}), but the more volatile \ce{NbCl5} precursor separated from the pellet during annealing and was not incorporated into the structure. 

The negligible ionic conductivity of this phase is consistent with design rules for multivalent ion conductors described in prior literature.
Rong et al. proposed that \ce{Mg^{2+}} mobility may be favorable in structures where \ce{Mg^{2+}} ions sit in energetically-disfavored sites (i.e., tetrahedra).\cite{rong2015materials} 
In \ce{MgZrCl6}, \ce{Mg^{2+}} occupies an octahedral coordination environment, which is more stable and thus less prone to hopping. 
Iton and See also noted that repulsive forces increase when a mobile ion moves through a site that is face-sharing with a site occupied by a highly-charged framework cation.\cite{iton2022multivalent}
The \ce{Zr^{4+}} sites in \ce{MgZrCl6} are face-sharing with 2/3 of the octahedral holes within the Van der Waals gap, inhibiting \ce{Mg^{2+}} mobility within that layer. 
Bond Valence Site Energy calculations suggest the lowest migration barrier in \ce{MgZrCl6} is 0.86~eV for interlayer hoping (Figure S8). 
This value is higher than the 0.6 eV cutoff used for prior theoretical work screening for \ce{Mg^{2+}} ion conductors.\cite{chen2019ionicMgCoatings} 
Our findings therefore further validate the design rules posed in these prior works.

Although \ce{MgZrCl6} proved not to be an \ce{Mg^{2+}} ion conductor, the synthetic approach may be useful for yielding phase-pure solids that include volatile precursors. In this case, \ce{ZrCl4} sublimes at 330~°C, but ball milling can trap it within the \ce{MgZrCl6} framework, allowing for subsequent heat treatment at 350~°C to crystallize the ternary without vapor loss. Many other chlorides also have low boiling or sublimation points, such as \ce{TiCl4} (boils at 136~°C), \ce{AlCl3} (sublimes at 180~°C), and \ce{NbCl5} (boils at 247~°C). %\cite{crcHandbook_Physical} 
Similarly, many bromides and iodides boil, sublime, or decompose at relatively low temperatures. The ball milling approach here may enable the synthesis of other temperature-sensitive halides.

\section*{Author Contributions}
C.L.R.- Conceptualization, Formal Analysis, Investigation, Formal Analysis, Visualization, Writing – Original Draft. 
A.S.- Investigation, Visualization. 
S.C.- Formal Analysis.
A.P.- Investigation. 
L.B.- Investigation.
J.R.N- Formal Analysis.
A.E.M.- Formal Analysis, Funding Acquisition. 
All authors edited the manuscript.
% CREDIT options
    % Conceptualization
    % Data Curation
    % Formal Analysis
    % Funding Acquisition
    % Investigation
    % Methodology
    % Project Administration
    % Resources
    % Software
    % Supervision
    % Validation
    % Visualization
    % Writing – Original Draft
    % Writing – Review & Editing

\section*{Conflicts of interest}
There are no conflicts to declare.

\section*{Data availability}
Crystallographic data on \ce{MgZrCl6} are available at the CCDC (Deposition number 2421152).
Additional data supporting this article have been included as part of the Supporting Information.

\section*{Acknowledgments}
This work was authored at the National Renewable Energy Laboratory (NREL), operated by Alliance for Sustainable Energy, LLC, for the U.S. Department of Energy (DOE) under Contract no. DE-AC36-08GO28308. C.L.R. and A.E.M. acknowledge support from the Laboratory Directed Research and Development (LDRD) program at NREL. Thanks to Sita Dugu for helpful discussions of Raman spectroscopy. Use of the Advanced Photon Source at Argonne National Laboratory was supported by the U.S. Department of Energy, Office of Science, Office of Basic Energy Sciences, under Contract no. DE-AC02-06CH11357. Use of the Stanford Synchrotron Radiation Lightsource, SLAC National Accelerator Laboratory, is supported by the U.S. Department of Energy, Office of Science, Office of Basic Energy Sciences under Contract no. DE-AC02-76SF00515. The views expressed in the article do not necessarily represent the views of the DOE or the U.S. Government.

%%%% SUPP INFO in the same document, because cross referencing is a pain in the butt...
\section{Supporting Information for: Ball milling enables phase-pure synthesis of a temperature sensitive ternary chloride, \ce{MgZrCl6}}

\setcounter{page}{1}
\pagenumbering{roman}

\renewcommand{\thefigure}{S\arabic{figure}}
\setcounter{figure}{0}

\section{Methods}
\subsection{Synthesis}
The precursors and products are moisture sensitive and hydrolyze rapidly on exposure to air. Thus, all work was performed in an Ar glovebox or otherwise under inert atmosphere. 
\ce{MgCl2} (Sigma Aldrich, 99.99\%, AnhydroBeads\textsuperscript{\texttrademark}), and \ce{ZrCl4} (Thermofisher Scientific, 98\%, anhydrous) precursors were used as received and were mixed 1:1 mol ratios for the various reactions.
The ball milled (BM) mixture was prepared using a Fritsch Pulverisette 7 Premium ball mill with 45 mL zirconia jars and 5 mm diameter zirconia milling media (ca.~5 g sample mass, 80 g mass of milling media). 
Samples were milled for 50 cycles of 10 min at 500 rpm followed by a 2 min rest (10 h total milling time). 
The hand ground (HG) mixtures were prepared by grinding the reagents with a mortar and pestle for 10 minutes.
Heat treated (HT) samples of the HG and BM mixtures were prepared by first pelletizing the powders using a 6 mm die pressed at 10 MPa using a hydraulic press (ca. 400 mg per sample), then loading the pellets into quartz ampoules (10 mm ID, 12 mm OD) and flame sealing under vacuum ($<$30 mTorr, ca. 10 cm total sealed length). 
Sealed ampoules were then heated at 5~°C/min to the set point, dwelled for between 1 and 160 h, and allowed to cool naturally by turning off the furnace. 
Samples were then recovered in the glovebox, ground with an agate mortar and pestle, and prepared for subsequent analysis.

\subsection{X-ray Diffraction}
Laboratory X-ray diffraction was collected on a Rigaku Ultima IV with a Cu K$\alpha$ source. 
Samples were prepared on a zero-background silicon wafer and were covered with a polyimide tape to minimize exposure to oxygen and moisture. 
Select samples were characterized by high resolution synchrotron powder X-ray diffraction (SPXRD) via the mail-in program at the Stanford Synchrotron Radiation Light Source beamline 2-1 ($\lambda = 0.729487$~\AA{}).\cite{stone2023remote}
The SPXRD measurements were calibrated using a \ce{LaB6} standard. 
Rietveld analysis was conducted using Topas v6.
The \ce{MgZrCl6} structure was determined using the \ce{FeZrCl6} structure as a starting phase (ICSD Col. Code 39666)\cite{troyanov1992crystalFeZrCl6} and replacing the Fe with Mg.
Rietveld analysis was conducted by refining the lattice parameters, atomic positions, thermal parameters, and size broadening (Lorentzian).

\textit{In situ} SPXRD measurements were conducted at the 17-BM-B end station of the Advanced Photon Source at Argonne National Laboratory ($\lambda =  0.24101$~\AA). The PerkinElmer plate detector was positioned 700~mm away from the sample. 
Homogenized precursors were packed into quartz capillaries in an Ar glovebox and flame-sealed under vacuum ($<30$~mTorr). 
Capillaries were loaded into a flow-cell apparatus\cite{chupas2008versatile} and heated at 10~°C/min to 600~°C. 
Diffraction pattern images were collected every 30~s by summing 20 exposures of 0.5~s each, followed by 20~s of deadtime.
Images collected from the plate detector were radially integrated using GSAS-II and calibrated using a silicon standard.

\subsection{Pair Distribution Function Analysis}
Total scattering data used for pair distribution function (PDF) analysis was collected at room temperature on beamline 28-ID-1 at the National Synchrotron Light Source II ($\lambda$ = 0.1665~Å). Powder samples were packed in quartz capillaries (1.1~mm outer diameter, 0.9~mm inner diameter) and flame-sealed. The data was collected using a sample-to-detector distance of $\approx$217.7965~mm, yielding a usable $Q_\textrm{max} = 26$~Å$^{-1}$, and wide-angle X-ray PerkinElmer image plate detector ($2048 \times 2048$ array, 200~$\mu$m pixel pitch). 
For each sample, two images were collected and averaged where total sample exposure time was 60~s, detector exposure time was 0.5~s, and sleep time was 60~s. The detector position and alignment was calibrated using a \ce{CeO2} (SRM674b) standard. The data was normalized into $(Q)$ and transformed into the PDF $G(r)$ according to, $G(r) = \frac{2}{\pi}\int_{Q_{min}}^{Q_{max}}Q[S(Q)-1]\sin(Q(r))dQ$, utilizing PDFgetX3.\cite{juhas2013pdfgetx3} %(CITE: https://doi.org/10.1107/S0021889813005190) 
Experimental PDFs were modeled with structures derived from the crystal structure using {\sc PDFgui}.\cite{Farrow_2007} Instrumental parameters, $Q_\text{broad} =$ 0.02011~\r{A}$^{-1}$ and $Q_\text{damp} =$ 0.03506~\r{A}$^{-1}$, were obtained by modeling the PDF of Ni measured during the same experiment.  

\subsection{Raman Spectroscopy}
Raman spectra were collected on a Thermo Scientific Nicolet iS50 FTIR spectrometer with an FT-Raman module. A CaF$_2$ beamsplitter and InGaAs detector were employed for all reported spectra. 
Samples were analyzed with 0.15~W of laser intensity ($\lambda=1064$~nm), 300 replicate scans, autogain on, and a resolution of 2 cm$^{-1}$. An aperture setting of 1 was necessary to eliminate detector saturation for the HT sample, while a setting of 37 was employed for all other samples. Samples were prepared by placing powdered samples on a dual concavity microscope slide and sealing the concavity with polyimide tape. Raman spectra were collected by placing the prepared microscope slides in the appropriate sample holder with the polyimide tape facing down (away from the light source). A blank microscope slide was measured under identical conditions and used for background subtraction.

\clearpage
\section{XRD Analysis}

\begin{figure}[ht!]
    \centering
    \includegraphics[width = 3in]{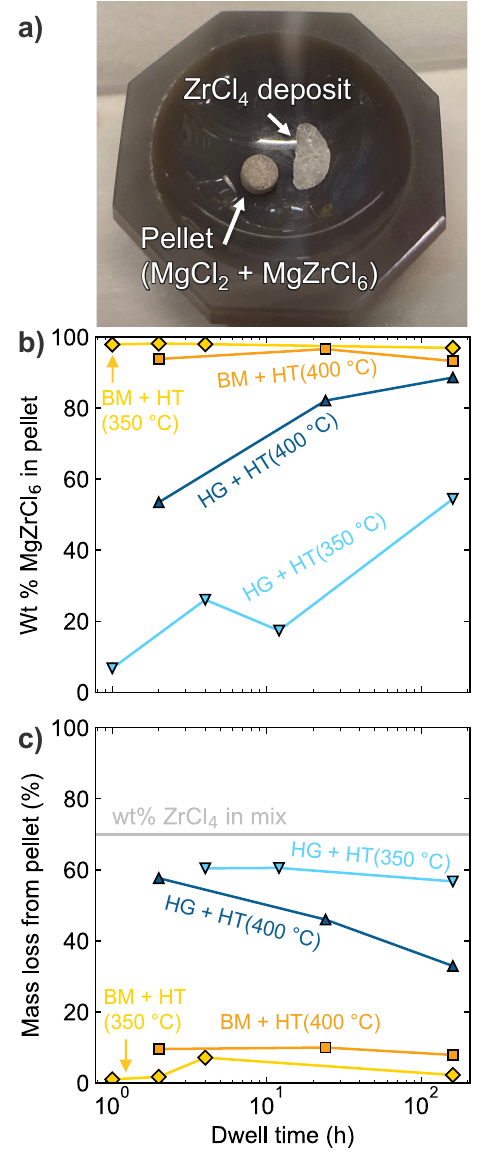}
    \caption{a) Photo of the pellet (containing \ce{MgZrCl6} and \ce{MgCl2}) and the \ce{ZrCl4} deposit recovered from a HG+HT reaction. b) Weight \% \ce{MgZrCl6} in the pellet determined using the Rietveld method (the remaining faction was \ce{MgCl2}). c) Mass loss from the pellet as a function of synthesis condition and dwell time, compared to the weight \% of \ce{ZrCl4} in the precursor mix. Representative XRD data and fits are shown in Figure \ref{fig:representative_XRD}.}
    \label{fig:mass_loss}
    %Source: 
\end{figure}
Ball milling (BM) was a key synthetic step because reactions initiated from hand-ground (HG) mixtures of \ce{MgCl2 + ZrCl4} did not produce phase-pure \ce{MgZrCl6} (Figure \ref{fig:mass_loss}). Instead, HG+HT reactions yielded partial formation of \ce{MgZrCl6}, along with unreacted \ce{MgCl2}. Some of the \ce{ZrCl4} transports as a vapor and deposits on the side of the ampoule (Figure \ref{fig:mass_loss}a). At temperatures above 350~°C, BM+HT samples also lose some \ce{ZrCl4}, indicating that the target phase decomposes partially. This suggests that the BM+HT and the HG+HT reactions converge towards an equilibrium as temperature increases: \ce{MgZrCl6 (s) <-> MgCl2 (s) + ZrCl4 (g)}. With increasing dwell time at 400~°C, the wt\% of \ce{MgZrCl6} in the pellet converges for the BM and HG (Figure \ref{fig:mass_loss}b) suggesting that the reaction in the HG mixtures is kinetically limited. However, atomic-scale mixing of BM precursors allows the \ce{MgZrCl6} structure to crystallize at low temperatures (and low \ce{ZrCl4} vapor pressure). 

\begin{figure}[ht!]
    \centering
    \includegraphics[width = \textwidth]{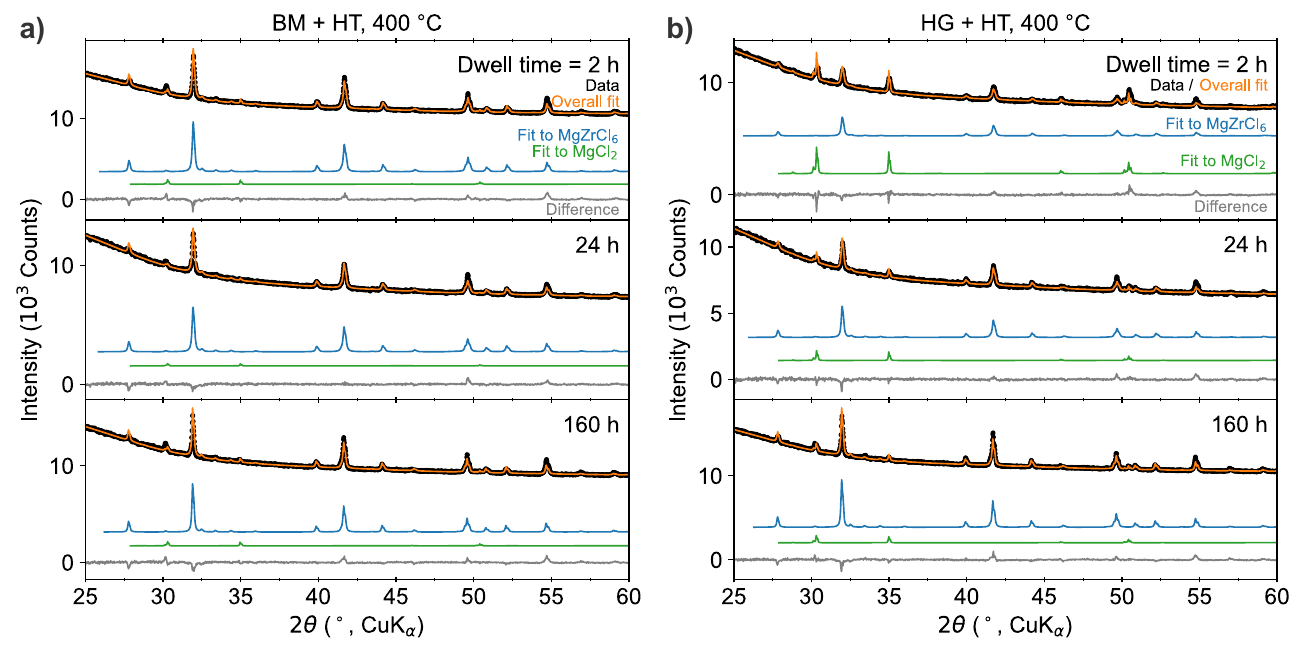}
    \caption{Representative Rietveld analysis of XRD following the 400 °C heat treatments of the a) ball milled vs. the b) hand ground precursor mixes.}
    \label{fig:representative_XRD}
    % Source: MgZrCl6_figureMapping.pptx and XRD ipynb
\end{figure}

Figure \ref{fig:representative_XRD} shows representative laboratory XRD patterns and corresponding fits. These fits were used to calculate the phase fractions shown in Figure \ref{fig:mass_loss}b. The BM+HT samples show consistently high phase fractions of  \ce{MgZrCl6} and low phase fractions of \ce{MgCl2}. In contrast, the HG+HT samples show a large \ce{MgCl2} fraction in the 2 h synthesis, with longer dwell times leading to less \ce{MgCl2} and more \ce{MgZrCl6}.

\clearpage
\section{Pair Distribution Function Analysis}

\begin{figure}[!h]
    \centering
    \includegraphics[width=\textwidth]{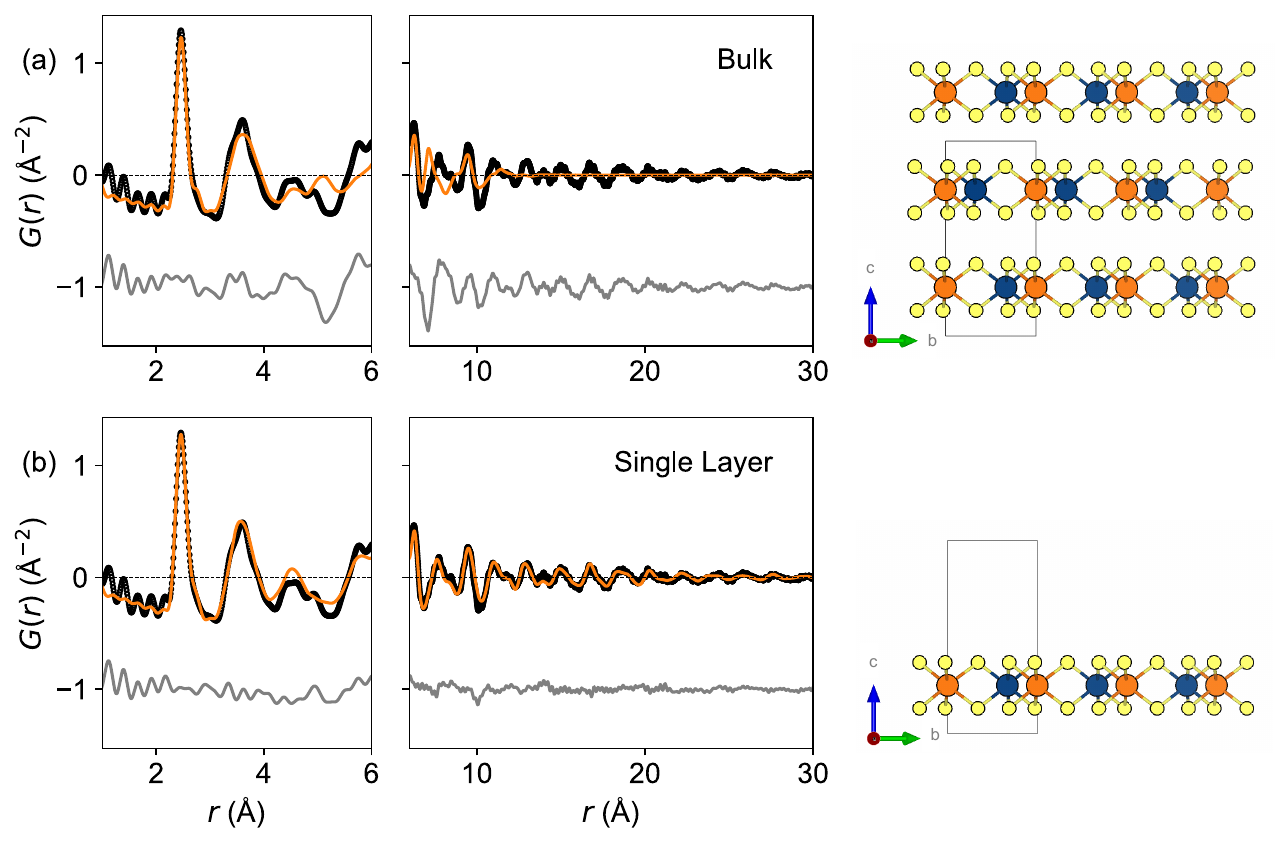}
    \caption{PDF of X-ray total scattering data from the BM sample of \ce{MgZrCl6} showing two different structural models: (a) the bulk crystal structure and (b) a single layer of \ce{MgZrCl6}, both with spherical truncation diameter. Data in black, fit in orange, difference in gray (offset vertically by 1 Å$^{-2}$).}
    \label{fig:PDFmodelcmpr}
\end{figure}

For the BM+HT sample, the bulk crystal structure produced a reasonable fit ($R_w$ = 20.4\%) for only refining a minimal number of parameters: the scale factor, correlated motion peak narrowing term ($\delta_1$ = 2.4(1)), the lattice parameters ($a$ = 6.35(1), $c$ = 11.84(3)), Cl fractional coordinates ($x$ = 0.34(1), $y$ =0.31(1), $z$ = 0.63(3)), and isotropic atomic displacement parameters for each element ($U_{Mg}$ = 0.023(3) \r{A}$^2$, $U_{Zr}$ = 0.012(6) \r{A}$^2$, $U_{Cl}$ = 0.026(9) \r{A}$^2$).  For the BM sample, two models were tested (Figure~\ref{fig:PDFmodelcmpr}).  The bulk model employed for the BM+HT sample captures part of the low $r$ pair correlations, but the model fails beyond $r\approx 4$~\r{A} ($R_w$ = 55.6\%), as illustrated in Figure~\ref{fig:PDFmodelcmpr}a.  The best simple model includes only a single layer of the \ce{MgZrCl6} crystal structure ($R_w$ = 25.8\%), as illustrated in Figure~\ref{fig:PDFmodelcmpr}b. Following the method of Chen, et al.,\cite{Chen:iu5034} a composite model was constructed using a single layer of the bulk crystal structure in a unit cell with a dramatically expanded $c$ axis and refining a minimal number of parameters (the scale factor, correlated motion peak narrowing term ($\delta_1$ = 2.2(2)), the lattice parameters ($a$ = 6.33(4), $c$ = 57.42, Cl fractional coordinates ($x$ = 0.36(1), $y$ =0.35(2), $z$ = 0.725), isotropic atomic displacement parameters for each element ($U_{Mg}$ = 0.044(84) \r{A}$^2$, $U_{Zr}$ = 0.030(28) \r{A}$^2$, $U_{Cl}$ = 0.03(1) \r{A}$^2$), and a spherical particle diameter (50(30) \r{A})).  The uncertainties of the $c$ axis and Cl $z$ parameters are undefined, as the unit cell is artificially elongated along the $c$ axis.  As the isolated single layer alone in an artificially large unit cell has the wrong atomic number density as the synthesized compound, the sloping baseline at low $r$ (e.g., $r<2$~\r{A}) and $r$-dependent envelope are not properly reflected with that phase alone.  Correcting for these aspects is easily implemented in {\sc PDFgui} following that of Ref.~\citenum{Chen:iu5034}; in summary, we include in the model a duplicated, linked second phase in the refinement with a drastically increased atomic displacement parameter to blur out the atomic correlations ($U_{iso}$ = 0.18(13) \r{A}$^2$, applied to all atoms) and a scale factor and a correlated motion parameter equal in magnitude but opposite in sign to the structural phase.  This ``composite'' model properly corrects the sloping baseline and shape envelope and is presented in Figures~\ref{fig:PDF}b and \ref{fig:PDFmodelcmpr}b.

\clearpage

\section{Raman Spectroscopy}
\begin{figure}[ht!]
    \centering
    \includegraphics[width = 0.7\textwidth]{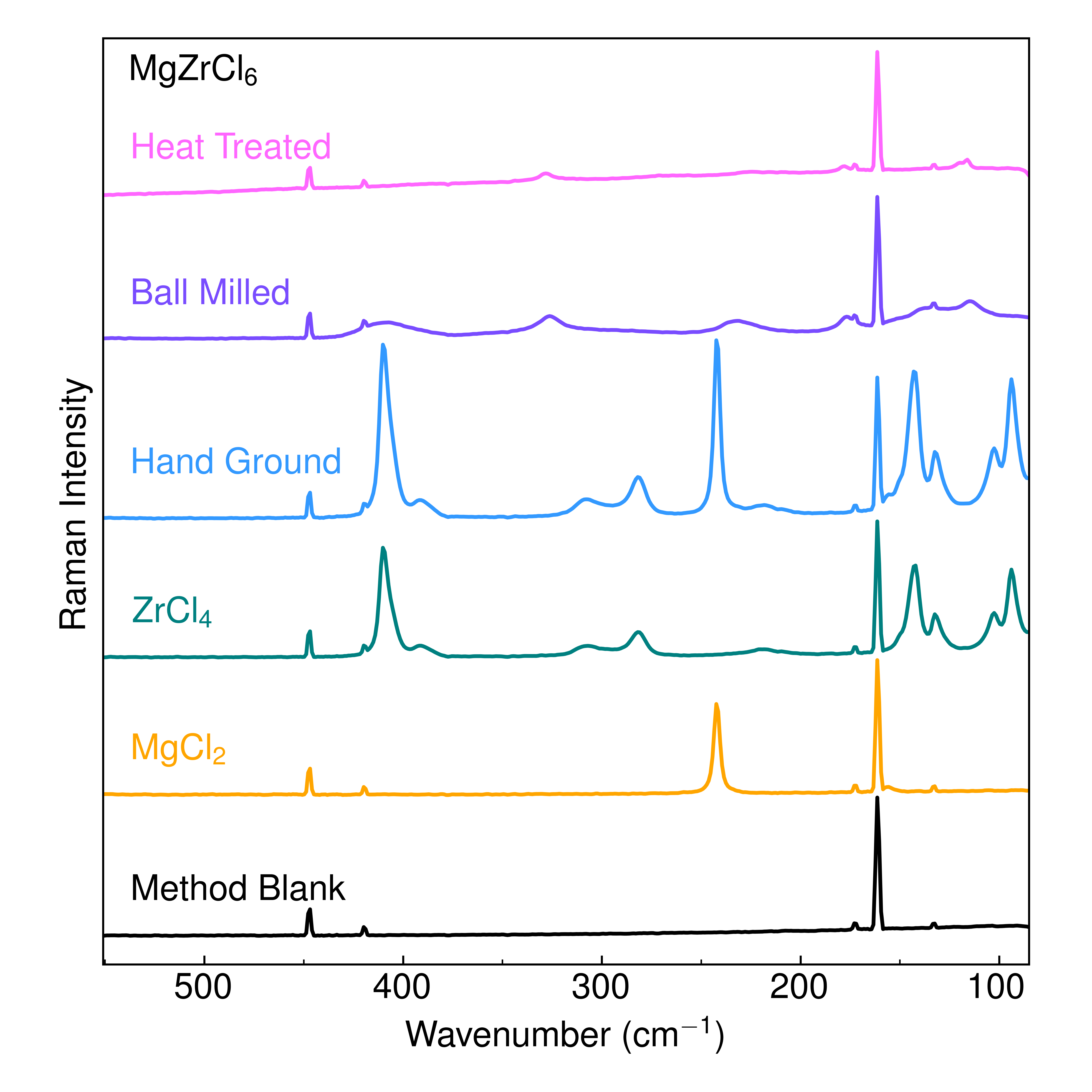}
    \caption{Raw Raman spectra of the samples shown in Figure \ref{fig:main_raman} prior to background subtraction of the blank sample holder (Method Blank).}
    \label{fig:raman_no_bkg_subtraction}
    % Source: austin
\end{figure}

\begin{figure}[ht!]
    \centering
    \includegraphics[width = \textwidth]{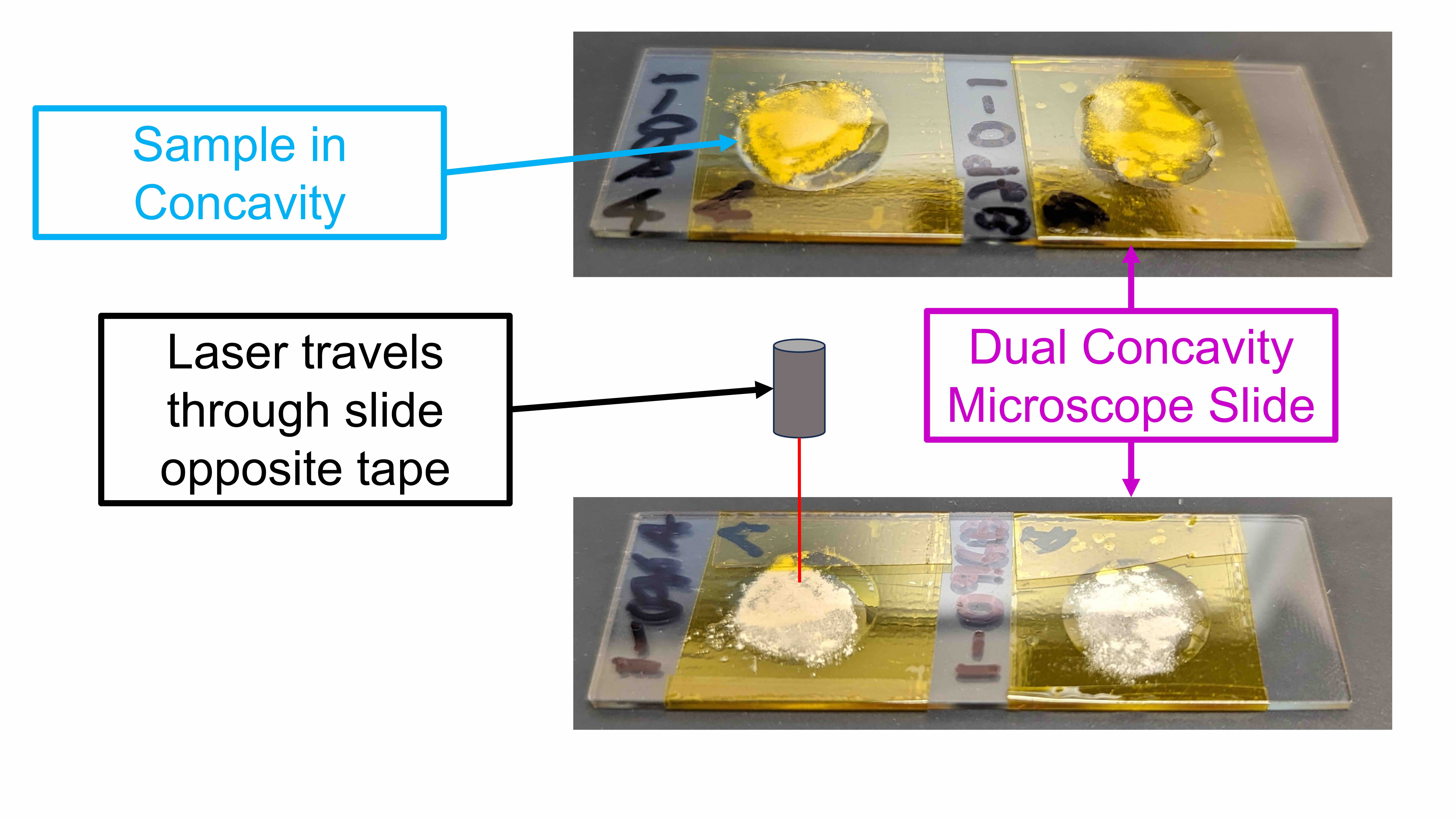}
    \caption{Diagram of the experimental setup for collecting Raman spectra.}
    \label{fig:raman_prep}
    %Source: austin
\end{figure}

The raw Raman spectra of the \ce{MgZrCl6} and precursor samples (Figure \ref{fig:raman_no_bkg_subtraction}) included substantial background signal from the sample holder (Figure \ref{fig:raman_prep}). Background subtraction was performed by subtracting the spectrum of the blank from the spectra of the samples, scaling the blank spectrum such that the large peaks from the background are no longer visible in the background-subtracted spectra. Background-subtracted Raman spectra are shown in the main body of the text (Figure \ref{fig:main_raman}).

\clearpage
\section{Electrochemical Impedance Spectroscopy}

\begin{figure}[!h]
    \centering
    \includegraphics[width = 3.2in]{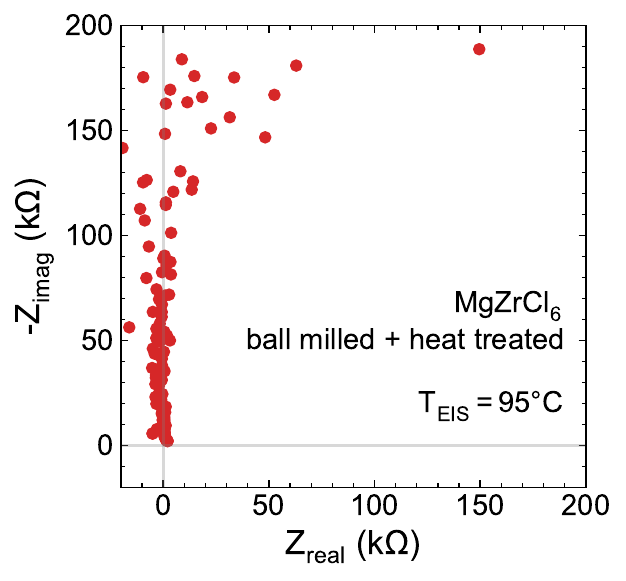}
    \caption{Nyquist plot for EIS of BM + HT \ce{MgZrCl6}, measured at 95~°C.}
    \label{fig:main_EIS}
    % Source: http://localhost:8888/notebooks/Documents/01_SSE/EIS/MgZrCl6%20eis.ipynb
\end{figure}

\begin{figure}[!h]
    \centering
    \includegraphics[width=0.7\linewidth]{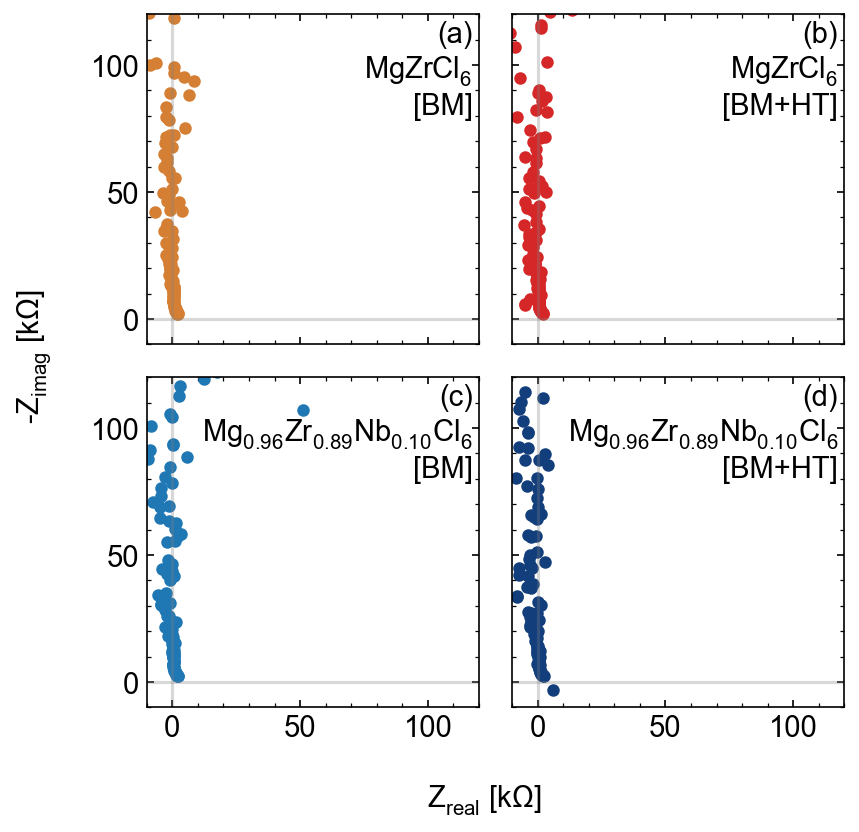}
    \caption{Nyquist plots for EIS of a) BM \ce{MgZrCl6} b) BM + HT \ce{MgZrCl6}, c) ball milled Nb-substituted \ce{MgZrCl6} and d) BM + HT sample of the Nb-substituted material. EIS was measured at 95~°C.}
    \label{fig:EIS_supp}
\end{figure}

Electrochemical impedance spectroscopy (EIS) measurements were performed to assess the possibility of \ce{Mg^{2+}} ionic conductivity. The \ce{MgZrCl6} did not exhibit measurable ionic conductivity, even at 95 °C (Figure \ref{fig:main_EIS}). Both the ball-milled nor crystalline \ce{MgZrCl6} phases exhibited behavior consistent with dielectrics (Figure \ref{fig:EIS_supp}a,b). Attempts to increase ionic conductivity via aliovalent substitution of \ce{Zr4+} with \ce{Nb^{5+}} similarly did not yield evidence of ionic conductivity (Figure \ref{fig:EIS_supp}c,d).

\clearpage
\section{Bond Valence Site Energy Calculations}

\begin{figure}[ht!]
    \centering
    \includegraphics{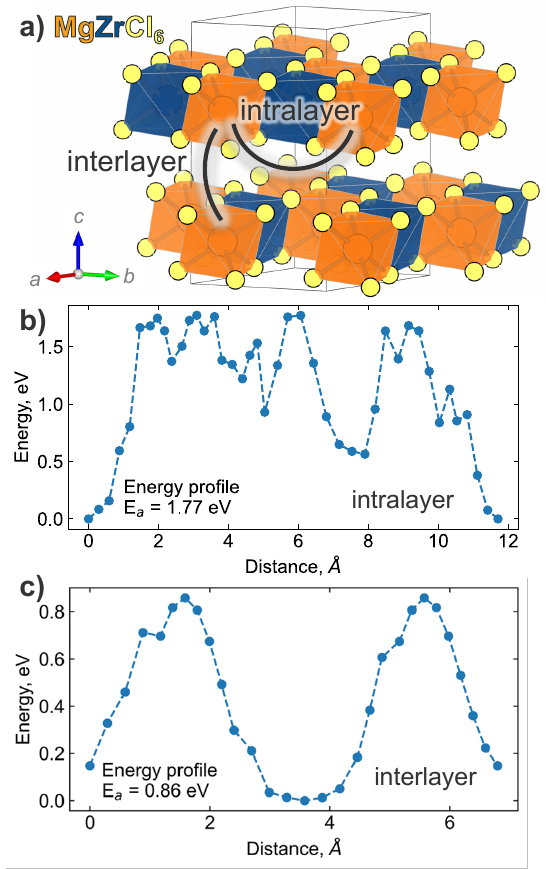}
    \caption{a) Structural model showing possible intralayer and interlayer migration pathways for \ce{Mg^{2+}} in \ce{MgZrCl6}. BVSE calculations for b) intralayer \ce{Mg^{2+}} migration show a pathway with a 1.77 eV barrier via inter-layer space and c) a lower energy pathway via interlayer migration with a 0.86 eV barrier. }
    \label{fig:BVSE_main}
    % Source: TODO
\end{figure}

We turned to Bond Valence Site Energy (BVSE) calculations to provide a potential explanation for the negligibly low ionic conductivity in MgZrCl$_6$. 
BVSE calculations suggest that the negligibly low ionic conductivity of \ce{MgZrCl6} may stem from high activation energy barriers (Figure \ref{fig:BVSE_main}). Intralayer migration barriers are calculated to be 1.77 eV. Interlayer migration has a lower barrier of 0.86 eV, but this value is still higher than the 0.6 eV cutoff used for prior theoretical work screening for \ce{Mg^{2+}} ion conductors.\cite{chen2019ionicMgCoatings} 
These high barriers are consistent with the prior design rules.\cite{rong2015materials, iton2022multivalent} 

\clearpage

\bibliography{MgZrCl6_bib}

\end{document}